\documentclass[iop]{emulateapj}
\usepackage{rotating}

\shorttitle{XUV and UV-B Disks in Low-Mass E/S0s}
\shortauthors{Moffett et al.}

\begin{document}

\title{Extended UV Disks and UV-Bright Disks in Low-Mass E/S0 Galaxies}

\author{Amanda J. Moffett \altaffilmark{1}, Sheila J. Kannappan \altaffilmark{1}, Andrew J. Baker \altaffilmark{2}, and Seppo Laine \altaffilmark{3}}

\altaffiltext{1}{Dept. of Physics \& Astronomy, University of North Carolina, Phillips Hall, CB 3255, Chapel Hill, NC 27599}

\altaffiltext{2}{Dept. of Physics \& Astronomy, Rutgers, the State University of New Jersey, 136 Frelinghuysen Rd., Piscataway, NJ 08854-8019}

\altaffiltext{3}{\emph{Spitzer} Science Center, California Institute of Technology, MS 220-6, Pasadena, CA 91125}

\begin{abstract}

We have identified 15 XUV (extended ultraviolet) disks in a largely
field sample of 38 E/S0 galaxies that have stellar masses primarily
below $\sim$$4 \times 10^{10}\, M_{\odot}$ and comparable numbers on
the red and blue sequences. We use a new purely quantitative XUV disk
definition designed with reference to the ``Type~1'' XUV disk
definition found in the literature, requiring UV extension relative to
a UV-defined star formation threshold radius. The 39$\pm$9\% XUV-disk
frequency for these E/S0s is roughly twice the $\sim$20\% reported for late-type
galaxies (although differences in XUV-disk criteria complicate the
comparison), possibly indicating that XUV disks are preferentially
associated with galaxies experiencing weak or inefficient star
formation. Consistent with this interpretation, we find that the XUV
disks in our sample do not correlate with enhanced outer-disk star
formation as traced by blue \emph{optical} outer-disk colors. However,
UV-Bright (UV-B) disk galaxies with blue UV colors outside their
optical 50\% light radii \emph{do} display enhanced optical outer-disk star
formation as well as enhanced atomic gas content. UV-B disks occur in
our E/S0s with a 42$^{+9}_{-8}$\% frequency and need not coincide with
XUV disks, thus their combined frequency is 61$\pm$9\%. For both XUV
and UV-B disks, UV colors typically imply $<$1 Gyr ages, and most such
disks extend beyond the optical $R_{25}$ radius. XUV disks occur over
the full sample mass range and on both the red and blue sequences,
suggesting an association with galaxy interactions or another
similarly general evolutionary process. In contrast, UV-B disks favor
the blue sequence and may also prefer low masses, perhaps reflecting
the onset of cold-mode gas accretion or another mass-dependent
evolutionary process. Virtually all blue E/S0s in the gas-rich regime
below stellar mass $M$$_{t}$~$\sim$~$5\times 10^{9}\,M_{\odot}$ (the
``gas-richness threshold mass'') display UV-B disks, supporting the
previously suggested association of this population with active disk growth.

\end{abstract}


\section{Introduction}

In hierarchical models of galaxy formation, galaxies often experience mergers that result in early-type remnants. Disk structures are also predicted to regrow around some of these remnants (e.g., \citealp{SM}; \citealp{Gv07}), allowing for transitions back from early- to late-type morphologies. Observationally, a transition stage from late to early types, brought about by interactions, may be glimpsed in the population of E+A (post-starburst) galaxies (e.g., \citealp{Y08}). However, observational evidence for the opposite predicted transition, that from early- to late-type morphology, has remained more elusive.

The ultraviolet regime offers a natural choice for studying possible disk growth. Recently, \emph{GALEX} has enabled the discovery of extended
ultraviolet (XUV) disks (e.g., \citealp{Th05}; \citealp{Gil05}). These XUV disks show ongoing star formation beyond the optical radii and
traditional star formation thresholds of late-type galaxies, providing an intriguing new look at galaxy disk growth in progress at
$z$$\sim$$0$. In a nearby galaxy sample
emphasizing late types, Thilker et al.\ (2007, hereafter T07) find a
20\% incidence of ``Type~1'' XUV disks, characterized primarily by
large radial extents and structured UV morphologies (versus ``Type~2''
XUV disks, which consist of less-extended UV-bright zones without
morphological specifications).

\emph{GALEX} has provided a useful platform for detection of star formation in early-type galaxies as
well. \citet{Kauf} find that extended UV emission is common in
high-mass bulge-dominated galaxies, likely associated with modest
reservoirs of cold gas in the disk that help fuel bulge and black hole
growth. Focusing specifically on galaxies with E/S0 morphology,
extended UV emission has also been seen in ring structures around
several S0 galaxies (\citealp{D09}; \citealp{CH}), and \citet{Th10}
recently identified an XUV disk around the nearby S0
NGC~404. \citet{SR10} have also identified several $z$$<$0.12
early-type galaxies with extended UV structures in far-ultraviolet
$HST$ imaging.

The presence of XUV disks, however, can have a variety of
interpretations. T07, for example, suggest an association of XUV disks
with interactions or minor perturbations. The raw
material for XUV disk formation could be acquired externally from such
interactions or from fresh cosmic gas accretion, either of which may
be consistent with the extended disks and rings of HI commonly observed around
E/S0s (e.g., \citealp{SW06}; \citealp{M06}; \citealp{Os07};
\citealp{Os10}). Another possibility for creating extended disks in
early types is the fallback of tidal tails in late stage mergers
(e.g., \citealp{H95}; \citealp{B02}; \citealp{N06}).

The evolutionary significance of disk growth may be greater in some of these scenarios than others. Of particular
interest is the scenario of cold mode gas accretion (e.g.,
\citealp{BD03}; \citealp{Ke05}; \citealp{DB06}; \citealp{Ke09}), which
may be linked to disk building in ``blue-sequence E/S0s,'' a recently
identified morphologically defined population of E/S0 galaxies on the
blue sequence in color versus stellar mass (\citealp{KGB}, hereafter
KGB). Blue-sequence E/S0s are primarily found in non-cluster environments (KGB), and as shown in KGB and \citet{We10a}, many display global gas reservoirs and specific star formation rates that
could allow the growth of significant new disks on relatively short
timescales.

Cold mode accretion occurs primarily below a critical shock heating stability mass (e.g., \citealp{BD03}; \citealp{Ke05}); this mass may coincide with an observed ``gas-richness threshold'' stellar mass at $M$$_{t}$ $\sim$ $5 \times 10^{9}\, M_{\odot}$, below which blue-sequence E/S0s become suddenly common, along with gas-dominated galaxies (\citealp{K04}; \citealp{KW}; see
KGB regarding corrected mass scale). This low-mass regime may be where the most active E/S0 disk growth occurs (KGB). Blue-sequence E/S0s also occur in modest numbers up to stellar masses of $\sim$$3 \times 10^{10}\, M_{\odot}$, the
bimodality mass of \citet{Kf03}, above which classical spheroids with older stellar populations begin to dominate.

To better understand the significance of recent disk star formation in E/S0s, we concentrate on the mass regime up to the bimodality mass and seek to quantify
the incidence of extended-disk star formation in a representative, largely field
sample of E/S0s. In \S 2, we introduce our
chosen sample and basic data. In \S 3, we discuss various methods for
identifying extended star formation, adopting the T07 Type 1 XUV-disk
designation as a reference. We then propose modifications to this
definition to create a purely quantitative classification that
reflects recent extended disk star formation in early
types. Since we are interested in the presence of disk star formation
in a general sense, we also introduce an alternative UV-Bright (UV-B) disk
definition, which can be used to identify significant disk star
formation not necessarily extended relative to traditional star
formation thresholds. In \S 4-5, we present demographics and properties
of our classified XUV and UV-B disks, and in \S 6 we
compare our results to various formation scenarios and results from the
literature. Finally, we provide a brief summary in \S 7.


\begin{figure}
\epsscale{1.}
\plotone{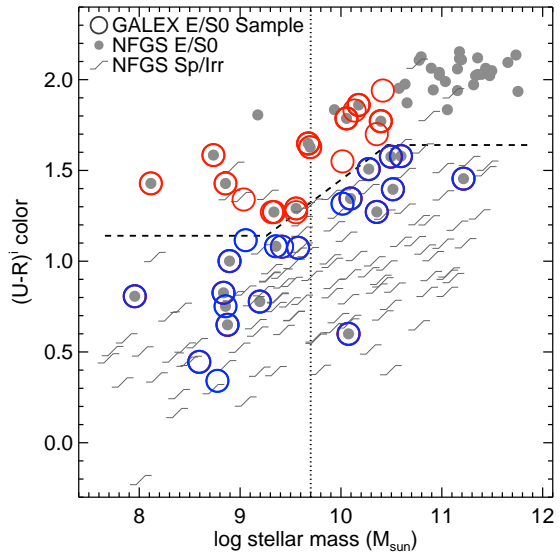}
\caption{\emph{GALEX} GI and archival E/S0 sample in color-stellar mass space. The small grey symbols indicate galaxies in the Nearby Field Galaxy
Survey, the parent sample for the majority of our E/S0s (\S 2). The
dashed line divides the red and blue sequences, and the vertical line
marks the gas-richness threshold mass (KGB). The 38 E/S0s with
\emph{GALEX} data are denoted by open circles.}
\label{RBfig}
\end{figure}

\section{Sample and Data Reduction}

Our ``\emph{GALEX} GI'' sample of 30 E/S0s was defined for
\emph{GALEX} program GI3-0046 and primarily draws on the Nearby Field
Galaxy Survey (NFGS, \citealp{Jan}). The sample was selected to
encompass all of the NFGS blue-sequence E/S0s and the majority of NFGS
red-sequence E/S0s in the stellar mass range below $\sim$$4 \times
10^{10}\, M_{\odot}$ (Fig.\ \ref{RBfig}), where many E/S0s have
substantial gas and settled blue-sequence E/S0s with the potential for
disk regrowth are observed (KGB). The NFGS provides a representative
sample of galaxies in the $z$$\sim$$0$ universe with a wide range of
luminosities, morphologies, and environments, allowing us to explore
the natural variety of stages in galaxy evolution. In addition to 25
NFGS E/S0s, the sample includes 5 blue-sequence E/S0s from the
``HyperLeda+'' sample of KGB with comparable archival data.

To augment this sample, we have cross-matched all
$M$$_{*}$~$\lesssim$~$4 \times 10^{10}\, M_{\odot}$ E/S0s in the
``HyperLeda+'' sample of KGB with the \emph{GALEX} and \emph{Spitzer}
archives to find sources imaged with exposure times similar to those
for our prior programs. Excluding Virgo Cluster members from this
cross-matched sample (consistent with the NFGS selection criteria), we
find eight additional E/S0s for our ``archival'' sample.

\begin{figure*}[t]
\epsscale{0.9}
\plotone{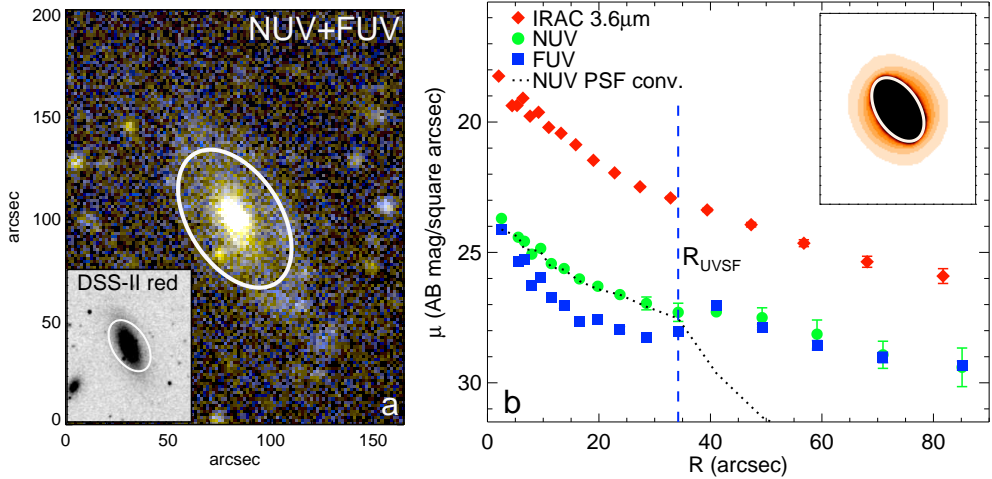}
\caption{Images and surface brightness profiles of NGC~4117, one of
several XUV-disk galaxies identified on the red
sequence. (a) \emph{GALEX} NUV+FUV color composite with
overlay of the NUV-derived star formation threshold, $R_{\rm UVSF}$ (see \S 3). The inset shows
the DSS-II red image with the same overlay for scale. (b) \emph{GALEX} and \emph{Spitzer} surface brightness
profiles. A vertical line marks $R_{\rm UVSF}$. The
black dotted line represents a profile extracted from the 2D
(re)convolution of the NUV PSF with the NUV galaxy light within
$R_{\rm UVSF}$ (see \S 3). The inset shows the (re)convolved image.}
\label{Galfig}
\end{figure*}

Our primary data are \emph{GALEX} NUV and FUV images at least as deep
as those of the Medium Imaging Survey (MIS). For comparison of UV and
optical morphologies, we employ DSS-II red images
(http://archive.stsci.edu/dss/). For profile analysis, we compare to
\emph{Spitzer} IRAC 3.6 $\mu$m imaging mostly obtained for program GO-30406
with typical exposure times of 480~s (although several archival
sources have exposure times down to 120\,s). The 3.6 $\mu$m imaging
serves as a proxy for $K$-band data, assuming the \citet{Le08}
conversion $I_{3.6}=0.55I_{K}$ (MJy~ster$^{-1}$). We use the notation $K_{80}$ to denote the 80\% light radius calculated using the 3.6 $\mu$m data, to indicate the direct analogy with the $K_{80}$ radius of T07.

We use \emph{GALEX} imaging in a pipeline-processed form with the zero
point calibrations of \citet{Mor07}. We apply foreground extinction
corrections based on \citet{S98} and \citet{C89}, but correction
factors for internal extinction are not applied (consistent with prior
XUV-disk studies). \emph{Spitzer} IRAC 3.6 $\mu$m imaging is also
pipeline processed and calibrated according to procedures outlined in
the IRAC Instrument Handbook
(http://irsa.ipac.caltech.edu/data/SPITZER/docs/irac/ iracinstrumenthandbook/). In addition to the pipeline processing, we apply a median
background subtraction procedure.

From these data, we extract radial surface brightness profiles and
magnitudes by totaling fluxes in elliptical
apertures. The parameters of these ellipses were determined from
isophotal fits to optical images (as reported in \citealp{Jan} for
NFGS galaxies) and newly derived using the IRAF ELLIPSE task and SDSS 
$g$-band imaging \citep{dr7} for non-NFGS sample galaxies (parameters for non-NFGS galaxies in the \emph{GALEX} GI sample from Stark et al., in prep). Detection and masking
of non-galaxy sources in these images was accomplished using
SExtractor \citep{Sxt}. For calculation of comparative \emph{GALEX}
and \emph{Spitzer} photometry, our UV and IR images were convolved
with an appropriately sized gaussian kernel to yield degraded images
with the same PSF FWHM as the lowest-resolution NUV images (FWHM $\sim$4.9$^{\prime\prime}$).

\begin{table*}

\begin{center}
\caption{Summary of Relevant UV-disk Definitions}
\label{Xdefs}
\begin{tabular}{ccccc}
\tableline\tableline 
\hline
\\[0.25pt]
Definition & Extent Criterion&Recent Star Formation Criterion\\
\noalign{\smallskip}\hline\noalign{\smallskip}
T07 Type 1 XUV & UV visually identified beyond $R_{\rm UVSF}$ & structured, bright UV with morphology different from optical \\
Purely Quantitative XUV & UV $>$3$\sigma$ above PSF shelf beyond $R_{\rm UVSF}$ &NUV$-K$ consistent with young population\\
UV-B & blue UV color beyond optical $R_{50}$ & NUV$-K$ consistent with $\gtrsim$10\% young population\\

\tableline
\end{tabular}
\end{center}
\end{table*}

\section{Identifying Extended Star Formation}
Here we discuss UV-based methods for identifying galaxies with recent
star formation in disks and extended disks. Ideally, we seek to
employ a purely quantitative method of classification. We also seek to answer two distinct questions about extended star formation in our
sample, for which different specific identification methods are relevant. First, does it occur beyond traditional star formation
thresholds? This question motivated the original ``Type~1'' XUV-disk
definition of T07, which we take as a reference in designing a purely
quantitative XUV disk definition (\S 3.2). Second, is it \emph{significant} (in a mass-contribution sense)
in the optical outer disks of galaxies? This question motivates our
introduction of a new ``Ultraviolet-Bright'' (UV-B) disk definition (\S
3.3; see also Table 1 for a summary of definitions used in this paper). 

We note that extension relative to UV-defined
star formation thresholds does not necessarily imply extension beyond the full optical extent of the galaxy. Thus, another
natural question about extended star formation is: does it extend beyond
the optical galaxy? We will treat the answer to this
question as a matter of investigation rather than definition, given
that the radial extent of star formation relative to the optical disk
may behave fundamentally differently in E/S0s vs. late-type galaxies,
for example, in the case of inside-out disk (re)growth.

\subsection{Prior Definitions}

A natural choice for answering our first guiding question, concerning 
star formation extended beyond traditional star formation
thresholds, is the T07 ``Type~1'' XUV-disk definition. T07 define Type~1 XUV disks as displaying more than one structured
UV-bright emission complex beyond a centralized surface-brightness
contour corresponding to the expected star formation threshold
(equated to an NUV surface brightness of 27.35 AB mag\,arcsec$^{-2}$
by T07, roughly matching typical H$\alpha$ and HI thresholds; we label
the corresponding radius $R_{\rm UVSF}$). In addition to extension
relative to this UV contour, the definition requires that the XUV
emission take on a different morphology from any underlying optical
emission. T07 also define a Type~2 XUV-disk classification, but this is
not geared towards tracing star formation beyond $R_{\rm UVSF}$, and an
issue\footnote[1]{The Type~2 XUV-disk classification requires
  FUV(AB)~$-$~$K$(AB) $\leq$ 4 in a large, optically low surface
  brightness zone within $R_{\rm UVSF}$ but outside $K_{80}$. Here ``large'' means an area at
  least seven times that enclosed within $K_{80}$. The Type~2
  definition was developed for a late-type sample and has proved
  problematic to apply to E/S0s, in that $R_{\rm UVSF}$ often lies
  inside the $K_{80}$ radius, or lies outside but not as far as the
  definition requires (see also \citealp{Me09} for further details).} with the definition
implies that we cannot apply it uniformly to early types. Thus, we do not
consider Type~2 XUVs further here and henceforth are referring to Type~1 XUVs when
we reference T07 XUV designations.

The T07 XUV definition is the basis for our new XUV definition (described in \S 3.2), but for completeness, we note that several other measures of bright and/or extended UV disks exist, most requiring high spatial resolution. For example, visual
classification of UV structures such as rings is common in the literature (e.g., \citealp{CH};
\citealp{SR10}; \citealp{Mar11}). A quantitative variant on extended UV disk identification involves measuring individual UV knots in the outer regions of galaxies (e.g.,
\citealp{ZC}). Another quantitative approach lacking the high resolution requirement is the blue integrated UV-color cut of \citet{Kauf}. However, with an integrated color cut alone the
correspondence between blue color and extended star
formation is not necessarily one-to-one. We modify this approach by adopting an \emph{outer-disk} UV color cut in our UV-B disk definition (see \S 3.3), addressing our second guiding question regarding significant star formation in the optical outer disks of galaxies.

\subsection{A New Purely Quantitative XUV Disk Definition}

To answer whether or not star formation occurs beyond
$R_{\rm UVSF}$, we adopt the T07 XUV-disk
definition as a useful reference definition and construct a purely quantitative alternative. Table 2 indicates the distribution and properties of the 16 XUVs we identify by the original T07 definition; see Figure \ref{Galfig} for an example. The primary criteria of the T07 XUV-disk classification are UV
extension relative to $R_{\rm UVSF}$ and association of this emission with recent star formation. In the following sections, we
discuss issues with these criteria that motivate elements
of our modified definition, including consideration of possible UV
upturn contributions and of the extended PSF shelf in the \emph{GALEX} NUV.

\subsubsection{Ensuring Young Ages}
A possible concern in identifying XUV disks in E/S0s is the prevalence
of the UV upturn, i.e., UV emission associated with old stellar
populations \citep{OC99}. To mitigate this issue, we identify XUV
disks in the NUV (in contrast to T07's use of a combination of FUV and
NUV data) since the UV upturn becomes stronger at FUV wavelengths. Nonetheless, 5 of the 16 XUVs we find using the original T07 XUV-disk definition have XUV-disk
region FUV$-$$K$ colors red enough to be consistent with a $>$1 Gyr
SSP (simple stellar population, as in T07 Figure 1). 

In general, the T07 requirement that UV emission take on a
different morphology from any underlying optical emission should
preclude classifying an underlying old population as a separate XUV
disk. However, the subjective requirement of structured emission can be difficult to apply
consistently to samples like our own: our galaxies tend to have smaller angular
sizes than those of T07, implying greater blurring at the
low angular resolution of \emph{GALEX}, so UV structure may be lost
or be difficult to assess. An XUV-disk definition relaxing this requirement of structured emission has recently been applied by \citet{Lem} to a sample containing both early and late types, and they experiment with using an FUV$-r$ cut to ensure young populations. 

\begin{figure}
\epsscale{1.}
\plotone{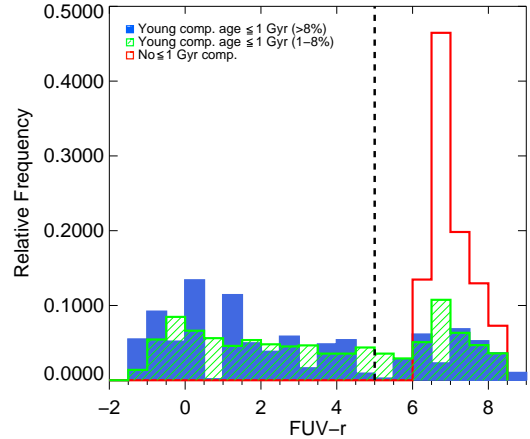}
\caption{FUV$-r$ color for selected composite stellar population models (grid as described in \S 3.2.1), illustrating issues with using this color as a clean young/old population divider. Blue and green histograms represent numbers of models with $>$8\% and 1-8\% young population contributions by mass, normalized to the total numbers of such models. The red histogram represents numbers of models containing no young (age $\le$ 1 Gyr) component, normalized to the total numbers of such models. The vertical dashed line indicates the color cut of \citet{Lem}, which appears to miss a significant fraction ($\sim$30\%) of the combined young model options that fall outside this cut with colors redder than FUV$-r$ $=$ 5.}
\label{Lemfig}
\end{figure}

Taking a similar approach but focusing on the NUV, we consider color cuts based on a suite of composite \citet{B03} stellar population models using a
Salpeter (1955) initial mass function, as described in KGB (see their
\S 2.3). These composite models are built from two (young and old)
components, with set age options, combined in a variety of ratios to
create a large model grid. Similar to the grid of KGB, the young SSP
age options are 5, 25, 100, 290, 640, and 1000 Myr, while the old SSP
age options are 1.4, 2.5, 3.5, ..., 13.5 Gyr. The young SSP
contributions can be 0\%, 1\%, 2\%, 4\%, 8\%, 16\%, 32\%, 64\%, or
100\% of the population mass. SSP metallicites allowed in the grid are
Z $=$ 0.008, 0.02, and 0.05. The young SSP can have 11 different
extinction values, but here we consider only zero-extinction models for comparison to observed outer-disk colors. We make no
explicit restriction on the metallicity combinations of the composite
population models we consider, although we find that
    consideration of metallicity restrictions that could be reasonable
    in specific circumstances, such as Z$_{\rm young}$ $\le$ Z$_{\rm
      old}$ or Z $\le$ Z$_{\rm solar}$, do not substantially change
    the model color distributions we report (see Figures \ref{Lemfig} and \ref{Nkfrac}).

In one version of their XUV-disk classifications, \citet{Lem} used a color cut at
FUV-$r$ $=$ 5, designed to separate galaxies with recent XUV-disk star
formation from those containing evolved populations (divider based on
empirical red/blue sequence division from \citealp{Wy07}). However, based on consideration of our
stellar population model grid (Fig.\ \ref{Lemfig}), this color selection can potentially
exclude up to $\sim$30\% of the composite populations with recently star-forming
components.

Thus, we search for a different color selection that better encompasses composite stellar populations with young components. As a result of the
aforementioned difficulties with using the FUV for this purpose and
the practical usefulness of making such a selection in bands where
data coverage is more complete, we prefer the NUV over the FUV. We find that NUV-based colors indeed display
a more cleanly defined region where populations are predominantly old (compare Figures \ref{Lemfig} and \ref{Nkfrac}). From the model
color distributions, it is apparent that the fraction of purely old
models increases significantly beyond NUV$-K$ $=$ 5, which is where
young model fractions start to decline as well. Thus, we choose to
exclude XUV disks with NUV$-K$ $>$ 5.

\subsubsection{Ensuring Extended Emission}
When applying the original T07 XUV-disk definition, classifiers must
subjectively identify the presence of extended emission beyond $R_{\rm
  UVSF}$. However, when classifying XUVs from \emph{GALEX} NUV
imaging, especially when considering galaxies with small angular
sizes, the $\sim$45$^{\prime\prime}$ shelf in the NUV PSF
(http://www.galex.caltech.edu/researcher/techdoc -ch5.html) may affect
this judgement. Thus, to design a quantitative test for extension
relative to $R_{\rm UVSF}$, we require that the NUV flux detected
\emph{outside} $R_{\rm UVSF}$ is significantly greater than
($>$3$\sigma$ above) the flux redistributed into this region by an
artificial second convolution of the NUV PSF with the flux
\emph{inside} this radius. This (re)convolution is in addition to the
natural convolution inherent in the images; see Figure \ref{Galfig}
for an illustration. We note that for the XUV disks we have identified
based on the original, subjective T07 definition, we have confirmed
that this requirement is always satisfied.

\subsubsection{Final Definition}

In summary, for our ``purely quantitative XUV-disk'' designation, we ensure UV emission beyond $R_{\rm UVSF}$
by requiring $>$3$\sigma$ emission above the NUV
PSF shelf, and we ensure recent star formation by requiring NUV$-K$ $<$ 5
in the XUV-disk region beyond $R_{\rm UVSF}$. Properties and demographics of these XUV disks are presented in \S 4 (see also Table 2)
and largely imply that this population is associated with
recent but not necessarily significant outer-disk
star formation. 

With our new definition, we identify a similar
fraction of XUV disks as when applying the traditional T07 Type~1 XUV-disk definition (see Table 2, Figure \ref{3pan}), but the overlap between these classifications is not perfect. Approximately 70\% of the traditionally identified XUVs are among XUVs identified with our purely quantitative method. In cases where the classifications do not overlap, the reason is either (1) insufficiently blue NUV$-K$ color to satisfy the new definition's color cut or (2) UV disk morphology not distinct enough from the optical to satisfy the T07 Type~1 definition. Our color cut is more conservative in rejecting XUV disks that may contain evolved populations than the T07 requirement of morphological differences compared to the optical. On the other hand, the T07 morphology requirement may recover XUVs with even weaker or more incipient star formation than our definition allows, where this star formation has not built up a detectable optical counterpart.

\begin{figure}
\epsscale{1.}
\plotone{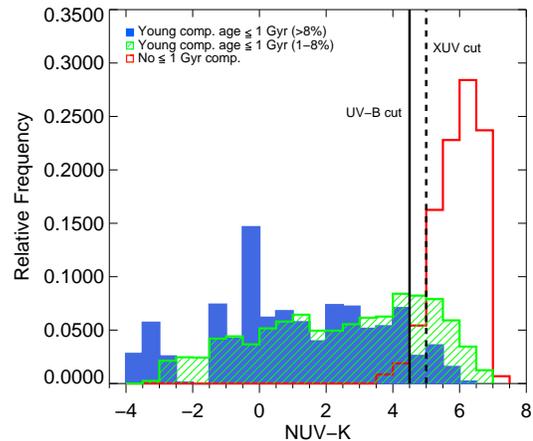}
\caption{NUV$-K$ color for selected composite stellar population models (grid as described in \S 3.2.1), illustrating NUV$-K$ cuts chosen for our analysis. Blue and green histograms represent numbers of models with $>$8\% and 1-8\% young population contributions by mass, normalized to the total numbers of such models. The red histogram represents numbers of models containing no young (age $\le$ 1 Gyr) component, normalized to the total numbers of such models. It is apparent that the fraction of purely old models increases significantly beyond NUV$-K$ $=$ 5, so we use this value to reject XUV disks likely to contain evolved populations as described in \S 3.2.1. A more conservative color cut at NUV$-K$ $=$ 4.5 appears necessary if we wish to select populations with a significant young population as in our UV-B disk classification (here $>$8\%, corresponding to the $\gtrsim$10\% requirement specified in \S 3.3).}
\label{Nkfrac}
\end{figure}

\subsection{UV-Bright (UV-B) Disk Definition}
To answer whether or not significant UV-detected star formation occurs in the optical outer-disk region, irrespective of extent beyond $R_{\rm UVSF}$, we construct a second quantitative classification.

We designate a population with a $\gtrsim$10\% young component by mass
as one containing ``significant'' star formation (in practice for our model set $>$8\%, \S 3.2.1). Considering the
aforementioned stellar population model grid,
a more conservative color cut than was used in the purely quantitative XUV
case appears necessary to select galaxies containing significant
recent star formation (Fig.\ \ref{Nkfrac}). Requiring NUV$-K$ $<$ 4.5 presents a natural
choice for this definition, given the falloff in the fraction of
models with a $\gtrsim$10\% young population component beyond this value.

To quantify our region of interest for this definition, i.e., the
optical outer disk, we select the region beyond the optical 50\% light
radius. Thus, our UV-B disk classification requires only NUV$-K$
$<$ 4.5 beyond the optical 50\% light radius. The properties and
demographics of the UV-B disks are presented in \S 5 (see also Table
2) from which we conclude that these galaxies correlate well with
enhanced optical disk star formation.

\begin{figure*}
\epsscale{1.}
\plotone{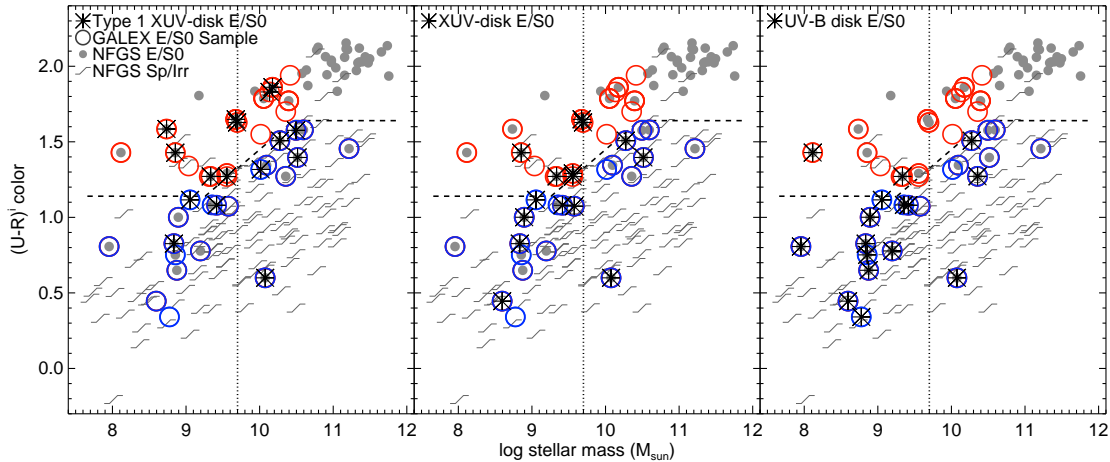}
\caption{\emph{GALEX} GI and archival E/S0 sample in color-stellar mass space (as in Figure 1), showing UV disks (asterisks) identified according to three different classifications: (left) the original T07 XUV-disk definition, (middle) our new, purely quantitative XUV definition, and (right) the UV-B disk definition.}
\label{3pan}
\end{figure*}

\newpage

\section{XUV Disk Properties and Demographics}
With our purely quantitative definition, we identify XUV disks in 15/38 or
39$^{+9}_{-9}$\% of our E/S0 sample (see Table 2 for the
identifications and Fig. \ref{allfig} for images of classified XUVs).
These XUV-disk classifications supersede the preliminary, purely
visual classifications of \citet{Me09}, which were made without
reference to $R_{\rm UVSF}$. In the following, we present
the demographics and basic properties of the identified XUVs.

\subsection{Extents and Ages}

The XUV disks in our E/S0s can extend beyond $R_{25}$, as has
been found in Type~1 XUV disks for late types (e.g., \citealp{Th05}; \citealp{Gil05}; T07;
see also \citealp{ZC}). We find radial extents (to the last measured
NUV point) $\sim$0.7$-$2.3$R_{25}$, with mean $\sim$1.3$R_{25}$ and
$\sim$70\% extending beyond $R_{25}$. Relative to the older
populations traced by near-IR light, the average radial extent of the
young XUV-disk component in our E/S0s is $\sim$2 times the $K_{80}$
radius. Relative to the centralized younger populations traced by NUV
light, the average radial extent of our XUV disks is $\sim$1.5 times
$R_{\rm UVSF}$.

Compared to XUV disks in late-type galaxies, our E/S0 XUV disks tend
to be redder. The reported outer-disk FUV$-$NUV colors of late-type
XUV-disk galaxies in the literature range primarily between small
negative values and $\sim$0.5 (\citealp{Th05}; Gil de Paz et al.\
2005, 2007). Our early-type XUV-disk galaxies have an average color of
$\sim$1.4 in the XUV-disk regions (similar to the early-type XUV-disk
galaxy NGC 404, \citealp{Th10}). However, the contour at $R_{\rm
UVSF}$ for our early-type XUVs tends to occur closer to $K_{80}$ than
it does for late-type XUVs (enclosing on average $\sim$3 times the
area of the $K_{80}$ contour versus $\sim$15 for late types; see
T07). Thus, redder XUV-disk colors in early types
may simply indicate a greater contribution from the underlying old
stellar population than is typical for late types.

The XUV-disk FUV$-$NUV colors we compute for our E/S0s are consistent
with $<$1 Gyr ages from simple stellar population models. We choose to report SSP-equivalent ages for our XUVs
 in light
of the inherent degeneracies involved in estimating separate old/young population ages from composite population models. We note that age estimates from stellar population models are affected by uncertainties in modeling the UV contribution from old stellar populations and will  also vary depending on the assumed star formation history. Comparing with \citet{B03} UV model colors for an instantaneous
starburst with $Z=0.02$ (as in T07 Figure 1), our average XUV-disk
FUV$-$NUV color of $\sim$1.4 corresponds to an SSP with an approximate age of 500 Myr. 

One of our XUV-disk galaxies does have XUV-disk region FUV$-$$K$
color red enough to be consistent with a $>$1 Gyr SSP (as in T07
Figure 1). However, all of our XUV-disk galaxies, including this red
FUV$-$$K$ case, display independent indicators for recent or potential star
formation, in the form of either H$\alpha$ or HI detections in the
NFGS or the literature.

\begin{figure*}
\epsscale{0.8}
\plotone{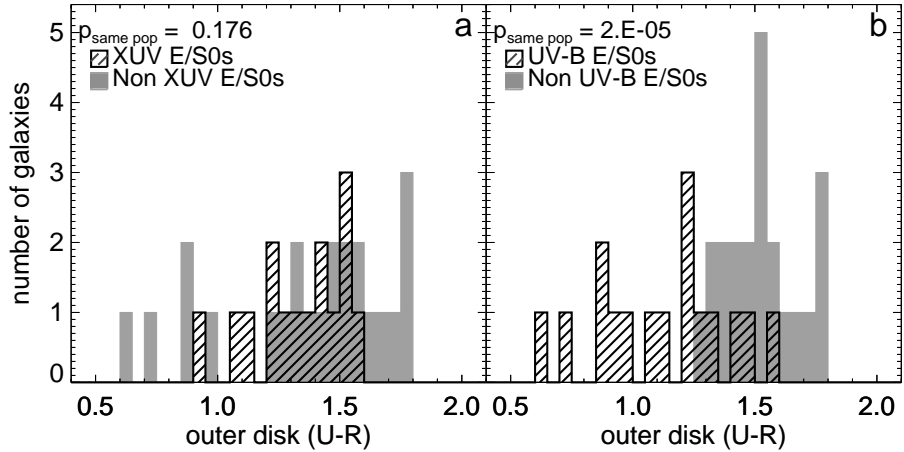}
\caption{Optical outer-disk colors for sample E/S0s, calculated between the 50\%--75\% $B$-band light radii for NFGS E/S0s and between the 50\%--75\% $g$-band light radii for all others ($u-r$ color is used as a proxy for $U-R$ for non-NFGS galaxies, with a shift to $U-R$ color as applied in KGB). (a) Comparison of E/S0s with and without XUV disks,
illustrating that XUV-disk E/S0s do not show bluer optical outer-disk colors than E/S0s without XUV disks, i.e., do not show enhanced outer-disk star formation. (b) Comparison of E/S0s with and without UV-B disks, illustrating that UV-B disk E/S0s \emph{do} show bluer optical outer-disk colors than E/S0s without UV-B disks, i.e., \emph{do} show clearly enhanced outer-disk star formation.}
\label{supfig}
\label{UVBcols}
\end{figure*}

\begin{figure}
\epsscale{1.}
\plotone{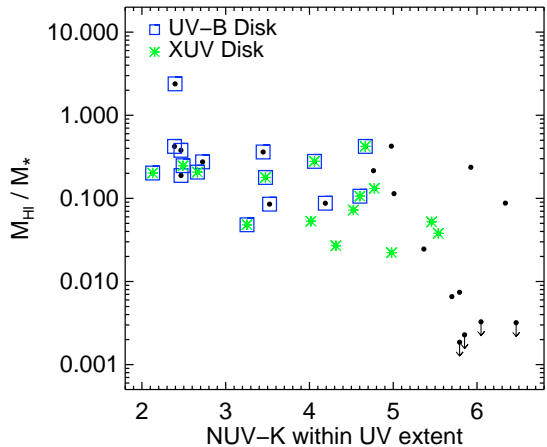}
\caption{HI content versus NUV$-K$ color measured within the detected NUV extent, illustrating the trend towards enhanced HI content in UV-B disk galaxies (blue squares). Black points represent all sample galaxies with HI data ($M_{\rm H\,I}$ from references noted in Table 2), and green stars represent XUV disks. Note that the plotted NUV$-K$ colors are not those used for classification of either XUV or UV-B disks.}
\label{HINK}
\end{figure}

\subsection{Demographics}
We find XUV disks in both red- and blue-sequence E/S0s and over a wide
range in stellar mass (Fig.\ \ref{3pan}). On the red sequence, the
XUV-disk frequencies are 0$^{+23}$\% and 60$^{+18}_{-20}$\% above and
below the gas-richness threshold mass (at stellar mass $M_{\rm t}$
$\sim$ $5 \times 10^{9}\, M_{\odot}$, KGB), respectively. On the blue
sequence, the corresponding frequencies are 33$^{+22}_{-18}$\% and
50$\pm$18\%.

If we ignore mass dependence, we find no clear evidence for a
preference in XUV-disk incidence between red- and blue-sequence
E/S0s. Assuming a probability for an XUV-disk galaxy to be on
the blue sequence equal to the overall sample blue-sequence fraction,
binomial statistics yields a 46\% probability of obtaining at least
the number of XUV-disk galaxies observed on the blue sequence out of
the total number of XUVs identified.

Likewise, if we ignore sequence
dependence, we find that the XUV-disk galaxy stellar mass distribution
is not significantly different from that of the parent E/S0 sample (61\%
probability of being drawn from the same distribution in a
Kolmogorov-Smirnov test). Binning the data in mass yields a hint of a
difference: the frequencies of XUV disks are 19$^{+15}_{-10}$\% and
55$^{+12}_{-13}$\%, respectively, above and below $M_{\rm t}$,
although the significance of this difference is not high ($\sim$1.8$\sigma$ confidence). 

We note that XUV disks identified according
to the original T07 definition have an even more uniform color/mass distribution (Fig.\ \ref{3pan}). Considering the slope of the color-stellar mass sequences, this difference is consistent with what one might expect as a consequence of our purely quantitative XUV definition excluding XUV disks with the reddest colors.

\subsection{Star Formation}
Although our XUV disks reflect recent star formation, we find
that they do not show substantial recent star formation as traced by blue \emph{optical} outer-disk colors
(Fig.\ \ref{supfig}). Likewise, the E/S0s with XUV disks do not show enhanced
atomic gas content relative to the E/S0s without XUV disks, instead yielding a 36\%
Kolmogorov-Smirnov test probability of the same $M_{\rm
  H\,I}$/$M$$_{*}$ distribution (Fig.\ \ref{HINK}). These possibly counterintuitive results imply that XUV disks are not necessarily associated with strong star formation and may instead be associated with weak/incipient star formation due to a process affecting the galaxy population broadly, an idea that we return to in \S 6.

We note that the T07 requirement of different UV-optical morphology may pick out weak star formation to an even greater degree than our purely quantitative approach, since the UV-optical morphology difference could imply that the UV-detected star formation is not substantial or sustained enough to have built up an optical counterpart. 

\section{UV-B Disk Properties and Demographics}
\begin{figure*}
\epsscale{1.}
\plotone{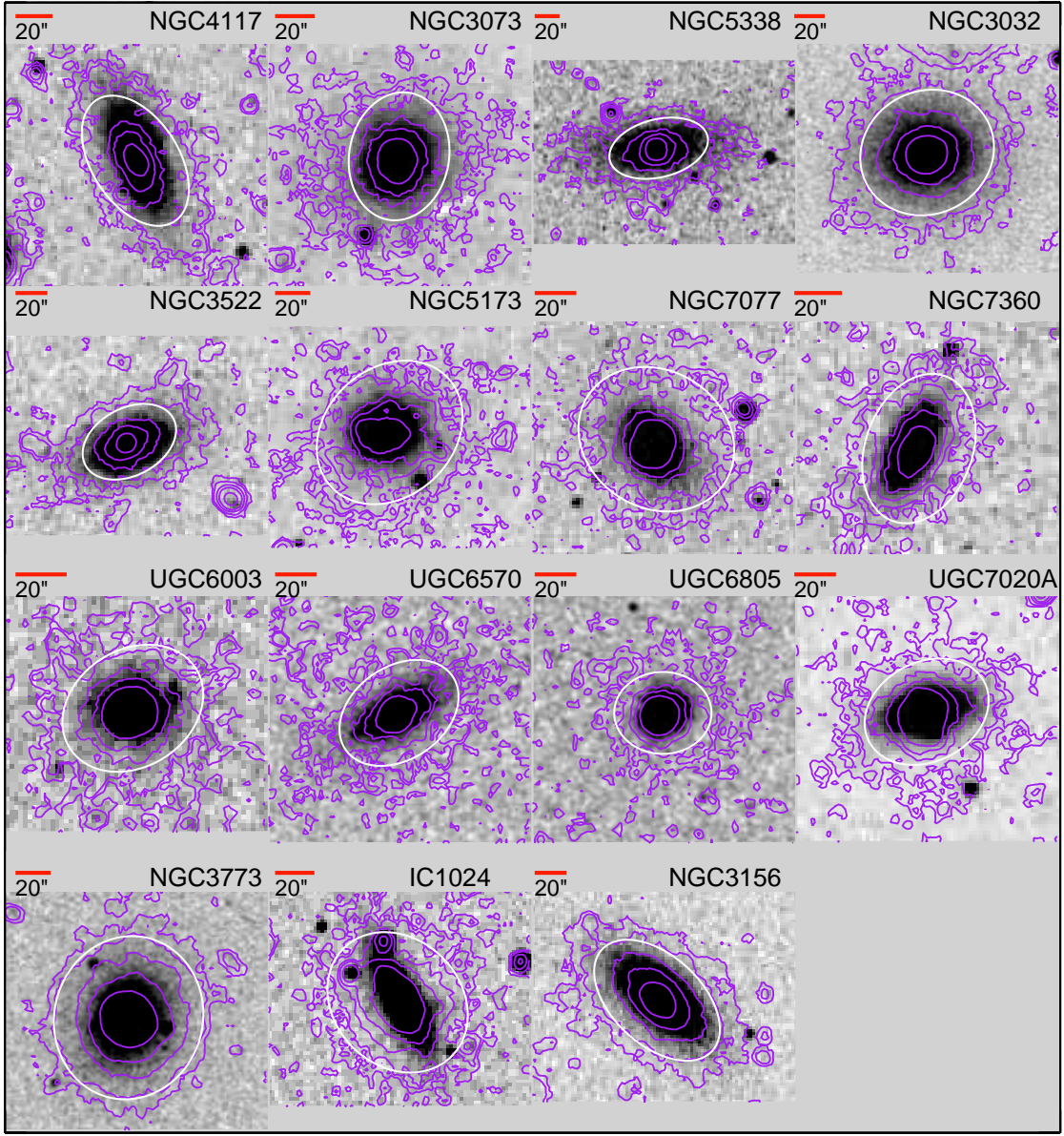}
\caption{Smoothed NUV contours (purple) overlaid on DSS-II red images of XUV-disk E/S0s. $R_{\rm UVSF}$ is indicated in white, and contours start at $\sim$28.6 AB mag\,arcsec$^{-2}$ and go up
by 2$\times$, 5$\times$, 10$\times$, and 25$\times$ in intensity. Eleven of these are Type~1 XUVs by the T07 definition: NGC4117, NGC3073, NGC5338, NGC3522, NGC5173, NGC7077, NGC7360, UGC6003, UGC7020A, IC1024, and NGC3156. Eight of these are also UV-Bs: NGC3073, NGC5173, NGC7077, UGC6003, UGC6805, UGC7020A, NGC3773, and IC1024.}
\label{allfig}
\end{figure*}

\begin{figure*}
\epsscale{1.}
\plotone{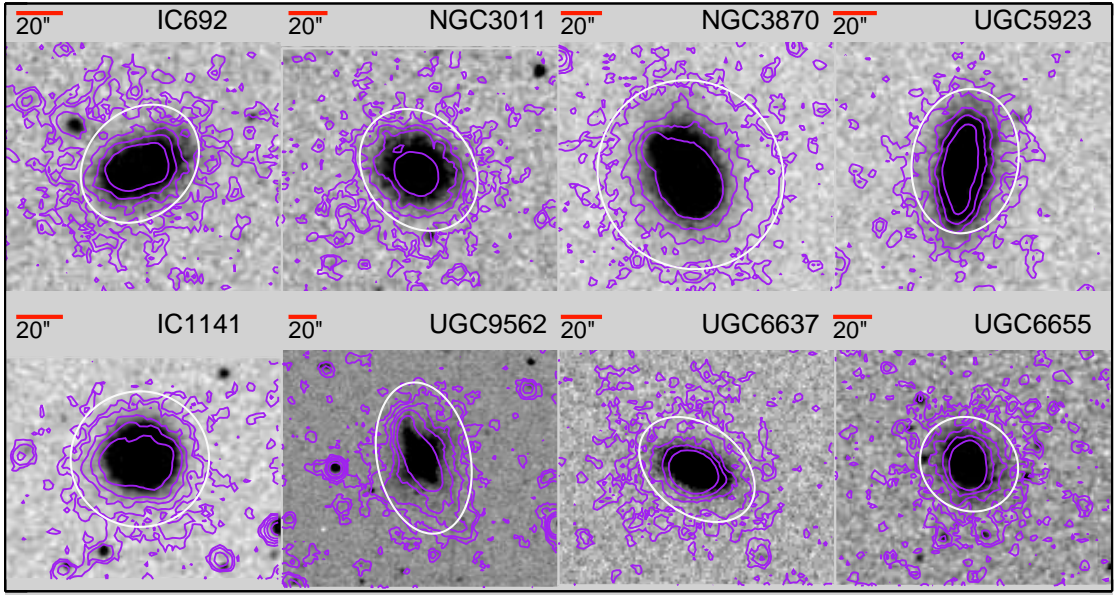}
\caption{Smoothed NUV contours (purple) overlaid on DSS-II red images of UV-B disk E/S0s that are not also XUV-disk E/S0s. $R_{\rm UVSF}$ is indicated in white, and contours start at $\sim$28.6 AB mag\,arcsec$^{-2}$ and go up
by 2$\times$, 5$\times$, 10$\times$, and 25$\times$ in intensity. None of these are Type~1 XUVs by the T07 definition. Some galaxies where the contours show extent beyond $R_{\rm UVSF}$ do not pass the test that this emission is $>$3$\sigma$ above the PSF shelf (see \S 3.2.2).}
\label{UVBfig}
\end{figure*}

With the UV-B disk classification, we identify 16/38 or
42$^{+9}_{-8}$\% of our sample as UV-B disks (see Table 2 for
identifications; Figs. \ref{allfig} and \ref{UVBfig} for images of
classified UV-Bs). Although we find similar frequencies of XUV and
UV-B disks in our sample, and about half of the galaxies
    with UV-B disks also host XUV disks, the overall properties and
demographics of these two classes display a number of differences, as
we discuss in the following sections.

\subsection{Extents and Ages}
Similar to our quantitatively identified XUV disks, the UV-B disks we
identify typically extend beyond $R_{25}$, with an average extent (to the last detected NUV point) of
$\sim$1.4$R_{25}$ and all extending beyond $R_{25}$. The average UV-B disk extent relative to the near IR
is slightly larger than for XUV disks at $\sim$2.3$K_{80}$ while the average extent
relative to the UV is smaller than for XUV disks at $\sim$1.3$R_{\rm
  UVSF}$.

The UV-B disk FUV$-$NUV colors we observe are also consistent with $<$1
Gyr SSP ages. For UV-B disk galaxies, the average FUV$-$NUV color outside the optical 50\%
light radius is $\sim$0.6, which corresponds to a
slightly younger $\sim$300 Myr SSP-equivalent age than is found for XUV disks. No UV-B disks display FUV$-$$K$ colors red enough to
imply SSP ages older than 1 Gyr.

\newpage

\subsection{Demographics}
In contrast to the widespread distribution of XUV disks, UV-B disks are preferentially found on the blue sequence and may prefer the low-mass regime as well (see color-mass
distribution in Fig.\ \ref{3pan}).

If we ignore mass dependence, we find clear evidence for a preference in
UV-B disk incidence between red- and blue-sequence E/S0s. Assuming a probability for a UV-B disk galaxy to be on the blue sequence
equal to the overall sample blue-sequence fraction, binomial
statistics yields a low 0.7\% probability of obtaining at least the
number of UV-B disks observed on the blue sequence out of the
total number of UV-Bs identified.

If we ignore sequence dependence, we
find that the UV-B disk galaxy and full sample stellar mass distributions
have an 8\% Kolmogorov-Smirnov test probability of being drawn from
the same distribution, which implies they are not conclusively distinct.
Similarly, the UV-B disk frequencies we calculate are 19$^{+15}_{-10}$\% and
59$^{+12}_{-13}$\%, above and below $M_{\rm t}$ respectively, which are more
different than in the XUV-disk case, but still only distinct at approximately
2$\sigma$ confidence.

\subsection{Star Formation}
In contrast to XUV disks, the UV-B disks in our sample \emph{do} correlate
with elevated star formation as traced by blue optical outer-disk
color (Fig.\ \ref{UVBcols}). E/S0s with UV-B disks also show enhanced HI content relative to E/S0s without UV-B disks (Fig.\ \ref{HINK}), with 0.1\% Kolmogorov-Smirnov test probability of the same $M_{\rm H\,I}$/$M$$_{*}$ distribution. Thus, it appears that UV-B disks are closely linked to significant star formation potential and pronounced optical outer-disk star formation.

\begin{figure*}
\epsscale{1.}
\plotone{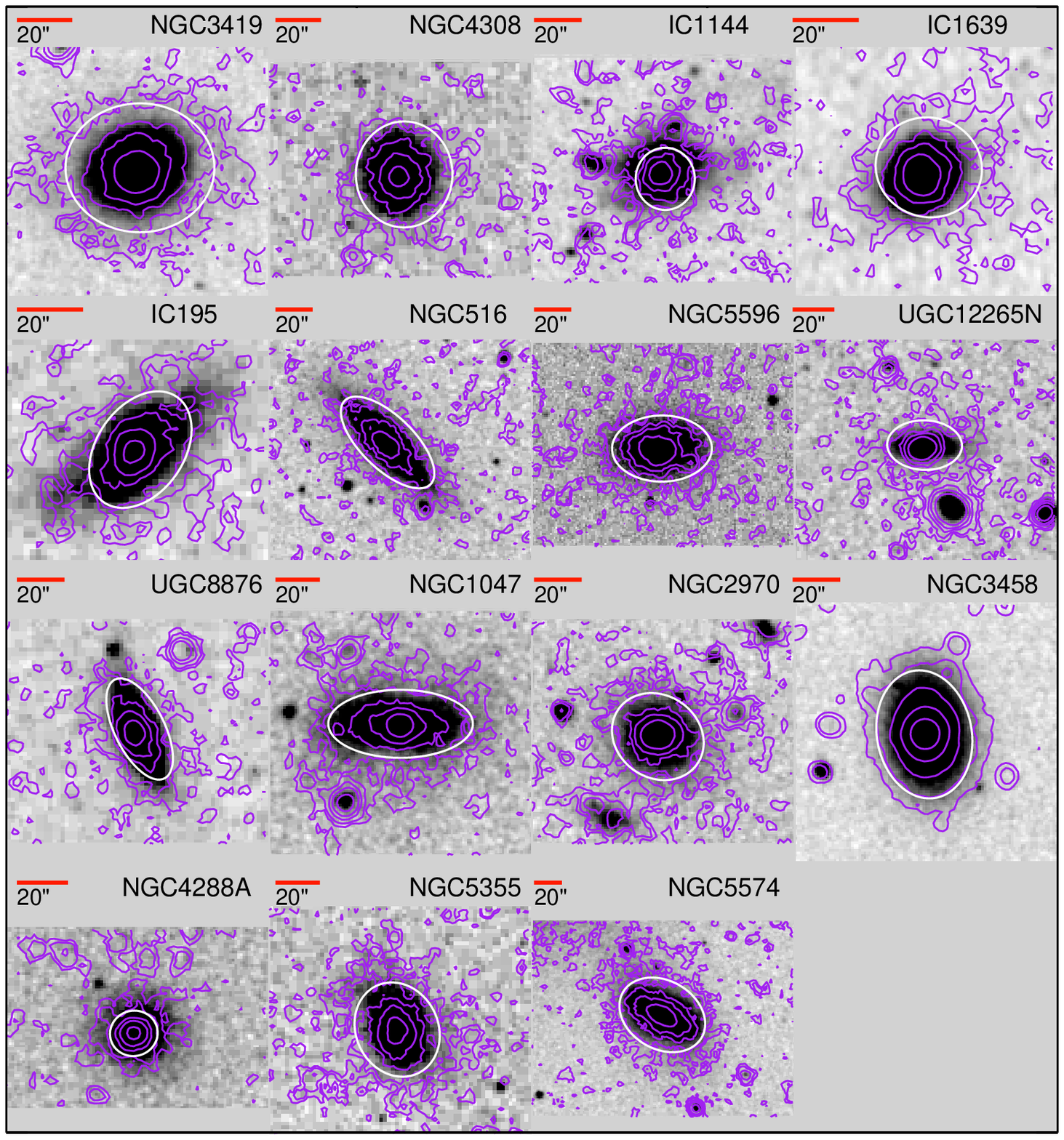}
\caption{Smoothed NUV contours (purple) overlaid on DSS-II red images of E/S0s in our sample without XUV or UV-B disks. $R_{\rm UVSF}$ is indicated in white, and contours start at $\sim$28.6 AB mag\,arcsec$^{-2}$ and go up
by 2$\times$, 5$\times$, 10$\times$, and 25$\times$ in intensity. Five of these are Type~1 XUVs by the T07 definition: NGC3419, NGC4308, IC195, UGC8876, and NGC5355. Some galaxies where the contours show extent beyond $R_{\rm UVSF}$ do not pass the test that this emission is $>$3$\sigma$ above the PSF shelf (see \S 3.2.2).}
\label{nonXfig}
\end{figure*}

\section{Discussion}
In this section, we compare our identified XUV- and UV-B disk galaxy
properties and demographics to XUV-disk and early-type galaxy formation
scenarios and related literature results. We note, however,
that uniform knowledge of the local and global environments of our
sample galaxies would be necessary to constrain these formation
scenarios and that uniform environmental data are not available for
our sample. Thus, study of the environmental properties of such
galaxies is deferred to future work.

\subsection{High Frequencies of XUV and UV-B Disks}
XUV and UV-B disks occur in our sample with individually
high, approximately 40\% frequencies, and a combined frequency of 61$\pm$9\%. Compared to classical ``red and dead'' expectations for early-type
galaxies, the high incidence of apparent extended star formation we
observe in XUV disks, with $\sim$70\% extending beyond the optical
$R_{25}$, is in itself a surprising result and may provide evidence
against E/S0 formation through quenching processes in the
low-mass, largely field regime we sample. Moreover, that we observe a
similarly high incidence of UV-B disks, which seem to relate more
closely to \emph{significant} star formation, and that \emph{all} extend past $R_{25}$ is even more remarkable. In addition, although
differences in samples and definitions complicate comparisons of
absolute XUV-disk frequencies, it is intriguing that we find a
frequency approximately twice the $\sim$20\% reported in late types by
T07 (see also \citealp{Lem}).

One possible explanation for the high incidence of XUV disks we
observe in E/S0s could be a formation channel that involves
mergers. Fallback of tidal tails in the late stages of a merger that is major
enough to produce a spheroid is a likely scenario for creating new
extended disks (e.g., \citealp{B02}; \citealp{N06}). Early type
galaxies at low luminosities/masses are largely ``fast rotators'' (as
per the \citealp{Em07} terminology), displaying disk-like dynamics reflecting the importance of gas in mergers related to their formation
(e.g., \citealp{D83}; \citealp{Em07}; KGB). Thus, if such mergers
often form XUV disks, the high frequency we observe in our mass regime
could be a natural consequence.

Another possible explanation for the high XUV-disk incidence in early
types compared to late types, related to the inferred weak nature of
XUV-disk star formation (\S 6.3), could be a bias due to the relative
ease of detecting small star formation events in E/S0s. Such events
may have a more detectable impact on the appearances/properties of
early types than late types, where they may be obscured by generally
higher levels of star formation.

\subsection{Ubiquity of XUVs Compared to UV-Bs}

The widespread distribution of the XUV disks in color and stellar mass seems to suggest an
association with evolutionary processes affecting the galaxy
population broadly. A potential scenario for creating extended,
star-forming disks around early types is external acquisition of
extended gas, whether delivered by companion interactions or fresh cosmic gas accretion, and subsequent conversion of this gas to
stars. In early-type galaxies, such extended disks or rings of HI are
frequently observed and often believed to be associated with external
accretion (e.g., \citealp{SW06}; \citealp{M06}; \citealp{Os07};
\citealp{Os10}).

T07 find that $\sim$75\% of their Type~1 XUV disks
show evidence for interactions or minor perturbations. An interaction scenario
could explain the widespread demographics of XUVs in our E/S0s, especially since the XUVs we identify using the original T07 definition, which favors discovery of the weakest XUVs, are the most broadly distributed (Fig. \ref{3pan}). To confirm such an association, more complete
knowledge of the companion statistics of our sample would be
needed.

For UV-B disks, which appear to have a somewhat
mass-dependent distribution, the higher UV-B disk frequency at low masses
(below $M_{\rm t}$) could hint at a gas delivery mechanism with a
preferred mass scale, as in the cold-accretion scenario (e.g.,
\citealp{BD03}; \citealp{Ke05}; \citealp{DB06}; \citealp{Ke09}). However, an in-depth
examination of the environments and group properties of a larger,
statistical sample of such galaxies will be necessary to
distinguish between various scenarios for producing extended star
formation in early types.

\subsection{Relationship to Star Formation and HI Content}
An important question to ask about the apparently young UV disks we
observe around E/S0s is: are they actually associated with substantial disk growth?

The higher ($\sim$40\%) frequency of XUVs in both red- and blue-sequence E/S0s vs.\ in late types seems to link XUV disks to a galaxy population associated with weak or inefficient star formation. Moreover, our E/S0 XUV disks do not show an association with blue optical outer-disk colors, nor with enhanced HI content (see \S 4.3). The association of XUV disks with weak or inefficient star formation is consistent with the
observation of a high ($\sim$70\%) rate of XUV disks in massive
optically low surface brightness galaxies, which are known for
inefficient star formation as well \citep{B08}. It is also consistent with
T07's result associating \emph{lower} SFR/$M_{\rm H\,I}$ with Type~1 XUV disks. In addition, simulations of XUV disks
in spiral galaxies show that star formation in these objects can
proceed for as long as 4 Gyr without producing enough stars to create
a high surface brightness optical component \citep{Bu08}.

In contrast, although XUV and UV-B classifications overlap, our UV-B disk galaxies as a class are characterized by bluer optical outer-disk colors and larger reservoirs
of HI gas than E/S0s without UV-B disks (see \S 4.4). Thus, UV-B disks are more closely associated with significant disk
star formation than are XUV disks. In addition, if we
consider the blue sequence below $M_{\rm t}$, where the numbers and properties of E/S0s suggest disk building is most active (KGB),
we find all sample galaxies save one are
classified as UV-B disks (Fig.\ \ref{3pan}). This strong link between UV-B disks 
and the sub-$M_{\rm t}$ blue sequence seems to support the scenario of
significant growth in the optical outer disks of E/S0s in this regime.

\section{Conclusions}
We have used UV, optical, and IR imaging to study extended-disk star
formation in a sample of 38 red- and blue-sequence E/S0s in the
stellar mass regime below $\sim$$4 \times 10^{10}\,
M_{\odot}$ and in primarily field environments. We introduce two new
classifications: a \emph{purely quantitative
version of the Extended Ultraviolet (XUV) disk classification}, akin to
the Type~1 XUV definition of Thilker et al.\ (2007; T07); and an \emph{Ultraviolet-Bright (UV-B) disk classification},
with NUV$-K$ color indicating $\gtrsim$10\% young population in the outer
optical disk beyond the 50\% light radius. We summarize key results from the application of these classifications below.

\begin{itemize}

\item We identify a high 61$\pm$9\% combined frequency of XUV and UV-B disks. Since the classifications partially overlap, this frequency reduces to separate 39$\pm$9\% and 42$^{+9}_{-8}$\% frequencies for XUV disks and UV-B disks, respectively. In the XUV-disk case, the observed frequency is approximately twice the $\sim$20\% reported by T07 for primarily late-type galaxies, although differences in XUV-disk criteria and possible detection biases could affect this comparison.

\item UV colors of both XUV and UV-B disks typically imply $<$1 Gyr ages, and most of the identified UV disks extend beyond the optical $R_{25}$ radius.

\item XUV-disk host
 galaxies occupy a widespread distribution in color and stellar mass, while UV-B disks more strongly prefer the blue sequence and may also prefer the low-mass regime.

\item XUV disks appear to be associated with \emph{low-level} star
  formation, whereas UV-B disks appear to be more clearly associated
  with \emph{significant} star formation. UV-B disk galaxies are also
  closely linked to the population of blue-sequence E/S0s in the
  stellar mass regime below the ``gas-richness threshold mass'' at
  $M$$_{t}$ $\sim$ $5 \times 10^{9}\, M_{\odot}$ (\citealp{KGB}; KGB), supporting the idea that such galaxies represent
  an actively disk-building population (KGB).

\end{itemize}

Our results suggest that XUV-disk formation could be related to a process that affects the galaxy population
  broadly, such as interactions, while UV-B disk formation could be related to a process with a mass-scale preference, such as cold-mode gas accretion. Existing data do not yet allow us to disentangle such effects in the E/S0 population, but the purely quantitative classifications we have developed in this work are well suited to application in larger statistical samples, which will allow us to construct a more complete picture of the local and global environments of star-forming E/S0s. In subsequent work, we plan to combine quantitative metrics of both disk building and environment in a large volume-limited survey in order to constrain the origin and future evolution of star-forming E/S0s and further probe the intriguing possibility of early-to-late-type transformation.

\acknowledgments

We thank S. Jogee for her role in acquiring the \emph{Spitzer} data, M. Haynes for the early release of \emph{GALEX} imaging of
NGC~3773, and the anonymous referee for suggestions that motivated substantial improvements to this work. We also thank D. Stark for helpful conversations on the topic of refining data analysis codes. We thank C. Clemens, K. Eckert, A. Leroy, M. Norris, and L. Wei for useful
discussions as well. AJM acknowledges support from the NASA Harriett G. Jenkins
Pre-doctoral Fellowship Program. This work uses observations made with
the NASA Galaxy Evolution Explorer. \emph{GALEX} is operated for NASA
by Caltech under NASA contract NAS5-98034. We acknowledge support from
the \emph{GALEX} Guest Investigator program under NASA grant
NNX07AT33G. This work uses observations made with the \emph{Spitzer}
Space Telescope, operated by the Jet Propulsion Laboratory, Caltech
under a contract with NASA. Support for this work was also provided by
NASA through an award issued by JPL/Caltech. This work uses observations from the SDSS; funding for the SDSS and SDSS-II has been provided by the Alfred P. Sloan Foundation, the Participating Institutions, the National Science Foundation, the U.S. Department of Energy, the National Aeronautics and Space Administration, the Japanese Monbukagakusho, the Max Planck Society, and the Higher Education Funding Council for England. The SDSS Web Site is http://www.sdss.org/.

{\it Facilities:} \facility{GALEX}, \facility{Spitzer}.

\begin{sidewaystable}
\centering
\scriptsize
\caption{\label{tbl1}}
\begin{tabular}{cccccccccccccc}
\tableline\tableline 
\hline
\\[0.25pt]
Galaxy &log($M$$_{*}$/$M$$_{\odot}$)& Seq & Morph. & Dist. & log($M_{\rm H\,I}$/M$_{\odot}$)& FUV$-$NUV& NUV$-$$K$& FUV$-$NUV & NUV$-$$K$  & T07 XUV? & XUV? &UV-B? & Sample \\
 & & & & (Mpc) & &($R > R_{\rm UVSF}$)
&($R > R_{\rm UVSF}$) &($R > R_{50}$) & ($R > R_{50}$)& & & \\
\noalign{\smallskip}\hline\noalign{\smallskip}
NGC3419&10.0&B&S0-S0/a&43.4&   9.1\tablenotemark{b}&2.9&5.0& 2.5  & 5.1 &Y&N&N& GI\\
                  NGC4117& 9.7&R&    S0& 19.0&  8.3&0.5&4.7& 0.7 &  5.1	&Y&Y&N&	GI\\
                  NGC3073& 9.1&B&  S0/a& 21.1&  8.5&1.6&4.5&  1.5  &4.4	&Y&Y&Y&	GI\\
               NGC4308& 8.7&R&    S0&  8.4&  $<$6.0&3.1&5.4&  2.3 & 5.6	&Y&N&N&	GI\\
               IC692& 8.9&B&         E&  21.4&8.4& 0.9& 2.9& 0.5  & 2.7	&N&N&Y&	GI\\
             NGC3011& 9.4&B&      S0/a&  25.7&8.3& 0.8& 3.3& 0.4 &  3.4	&N&N&Y&	GI\\
             NGC3870& 8.8&B&       Pec&  14.5&8.4& 0.4& 2.7& 0.3  & 2.7	&N&N&Y&	GI\\
             UGC5923& 8.1&R&      S0/a&   8.0&7.7& 1.1& 3.8& 0.7  & 3.6&N&N&Y&	GI\\
                  NGC5338& 8.9&R&    S0& 10.3&  7.3&2.0&4.5&  1.9 & 4.7	&Y&Y&N&	GI\\
              IC1141&10.4&B&      S0/a&  68.0&9.3& 0.7& 3.5&0.6   & 4.0 &N&N&Y&	GI\\
           IC1144&11.2&B&      S0/a& 175.4&$<$8.7& 2.3& 5.8& 2.3  & 6.0	&N&N&N&	GI\\
              IC1639&10.6&B&        cE&  76.0&8.4& 3.1& 5.3& 2.0 &  5.7 &N&N&N&	GI\\
                    IC195&10.5&B&  S0/a& 52.1&  9.4&1.9&6.2&1.9 &   6.2	&Y&N&N& GI\\
             UGC9562& 8.9&B&        S0&  25.2&9.3& 0.4& 2.5& 0.4  & 2.2	&N&N&Y&	GI\\
             NGC3032& 9.6&B&       Pec&  25.2&8.3& 2.0& 4.7&  1.8 & 4.8	&N&Y&N&	GI\\
                  NGC3522& 9.7&R&    S0& 22.9&  8.4&2.1&4.9&  2.0 & 5.1&Y&Y&N&	GI\\
           NGC516&10.1&R&        S0&  34.9&$<$7.4& 2.8& 5.6& 2.2 &  5.8&N&N&N& GI\\   
                  NGC5173&10.3&B&     E& 41.2&  9.3&0.8&3.9& 0.7 &   4.2&Y&Y&Y& GI\\ 
             NGC5596&10.4&R&        S0&  50.8&8.8& 1.2& 5.5& 0.6 &   5.2&N&N&N&	GI\\
                  NGC7077& 8.8&B&  S0/a& 18.9&  8.2&2.1&3.1& 0.7 &   3.0&Y&Y&Y&	GI\\
                  NGC7360&10.5&B&     E& 67.9&  9.6&0.6&4.4& 0.6  &  4.5&Y&Y&N&	GI\\
           UGC12265N&10.1&B&        S0&  82.8&9.4& 1.1& 4.7& 1.0  &  4.8&N&N&N&	GI\\
                  UGC6003&10.1&B&  S0/a& 84.2&  9.4&1.5&1.9& 0.8 &   2.7&Y&Y&Y&	GI\\
             UGC6570& 9.6&R&      S0/a&  28.6&8.4& 1.4& 4.5&  1.0 &  4.6&N&Y&N&	GI\\
             UGC6637& 9.2&B&        S0&  31.5&8.6& 0.8& 3.7& 0.3 &   2.9&N&N&Y&	GI\\
             UGC6655& 8.0&B&        S0&   8.8&7.2& 0.8& 3.4& 0.3 &   2.8&N&N&Y& GI\\  
             UGC6805& 8.9&B&        S0&  20.3&7.6& 1.3& 3.9& 0.6 &   3.5&N&Y&Y&	GI\\
                 UGC7020A& 9.3&R&    S0& 26.7&  8.6&0.7&3.5& 0.6 &   3.5&Y&Y&Y&	GI\\
                UGC8876&10.2&R&  S0/a& 36.7& $<$7.7&1.6&6.3&  1.5 &  6.4&Y&N&N&	GI\\
NGC3773& 8.6&B&       Pec&  10.5& 7.9&---\tablenotemark{c}&3.4&---\tablenotemark{c}&3.2&N&Y&Y&GI\\
IC1024& 9.4&B& S0Pec& 20.4& 9.0\tablenotemark{b}&0.8&4.0&   0.7 &   4.3           &Y&Y&Y& archival\\
NGC1047& 9.0&R&      S0/a&  19.1& 8.7\tablenotemark{b}& 2.3& 5.3&  2.5   &   5.2  &N&N&N&  archival\\
NGC2970& 9.3&R&      S0/a&  22.7& ---\tablenotemark{a}& 2.4& 5.2 & 2.0   &   5.2  &N&N&N&  archival\\
NGC3156& 9.6&R&    S0& 15.3& 7.9\tablenotemark{b}&2.6&4.9&  2.4  &    5.1         &Y&Y&N&  archival\\
NGC3458&10.4&R&        S0&  27.6& ---\tablenotemark{a}& 1.9& 6.2&   1.7  &   6.5 &N&N&N&  archival\\
NGC4288A&10.4&R&        S0&  100.8& ---\tablenotemark{a}& 1.4&6.1&  1.1   &  5.9 &N&N&N&  archival\\   
NGC5355&10.1&R&S0/a-S0Pec&34.4&9.5\tablenotemark{b}&---\tablenotemark{c}&5.7&---\tablenotemark{c}&5.8&Y&N&N& archival\\
NGC5574&10.0&R&      S0/a&  23.2& 7.9\tablenotemark{b}& 2.0& 5.7& 2.3 &    5.7    &N&N&N& archival\\

\tableline
\end{tabular}
\tablecomments{Derived FUV$-$NUV and NUV$-$$K$ colors outside $R_{\rm UVSF}$ and beyond the optical 50\% light radius (the XUV-disk and UV-B disk classification regions, respectively) plus ancillary data and classifications using the original T07 XUV-disk definition, our purely quantitative XUV-disk definition, and the UV-B disk definition. Ancillary data are from Jansen et al.\ (2000a,b), KGB, and \citet{We10a} except as noted. Distances assume $H_{0}$=70~km~s$^{-1}$~Mpc$^{-1}$. $^{\rm{a}}$ HI data unavailable.$~$ $^{\rm{b}}$ HI data from HyperLeda \citep{Pat}. $^{\rm{c}}$ No FUV imaging available.}

\end{sidewaystable}

\end{document}